\documentclass[twocolumn,showpacs,preprintnumbers,amsmath,amssymb]{revtex4}

\usepackage{dcolumn}
\usepackage{bm}
\usepackage[english]{babel}
\usepackage[latin1]{inputenc}
\usepackage{amsmath}
\usepackage{amsfonts}
\usepackage{amssymb}
\usepackage{graphicx}
\usepackage{pstricks}
\usepackage{pst-node}
\usepackage{wrapfig}
\usepackage{epsfig}
\usepackage{amsbsy}
\usepackage{rotating}
\begin{document}

\preprint{APS/123-QED}

\title{Single cell mechanics: stress stiffening and kinematic hardening}
\author{Pablo Fern\'andez}
\altaffiliation{Present address: Lehrstuhl f\"ur Biophysik E22,
Technische Universit\"at M\"unchen, James Franck Stra\ss{}e, D-85748 Garching,
Germany}
\author{Albrecht Ott}
\altaffiliation{Present address: Biologische Experimentalphysik, Universit\"at des Saarlandes,
D-66041 Saarbr\"ucken, Germany}
\affiliation{Experimentalphysik I, Physikalisches Institut, 
Universit\"at Bayreuth,
D-95440 Bayreuth, Germany}

\date{\today}

\begin{abstract}
Cell mechanical properties are fundamental to the organism but remain poorly understood.
We report a comprehensive phenomenological framework for the nonlinear rheology of single fibroblast cells:
 a superposition of elastic stiffening and viscoplastic kinematic hardening.
Our results show, that in spite of cell complexity its mechanical properties
can be cast into simple, well-defined rules, which provide mechanical cell strength and robustness via control of crosslink slippage.
\end{abstract}

\pacs{87.15.La, 83.60.Df,  83.60.La, 87.16.Ka}
\maketitle

Intracellular transport, cell locomotion, resistance to external mechanical stress and
other vital biomechanical functions 
of eukaryotic cells are governed by the cytoskeleton, an active biopolymer gel 
\cite{braybook}.
This gel is made of three types of biopolymers, actin, microtubules and intermediate filaments, crosslinked by a multitude of proteins with different properties in terms of connection angles, bond strengths and bond lifetimes. 
The actin cytoskeleton --the major force-sustaining structure
in our experiments--  is made of filaments of about micrometer length and presents a weak local structural order. 
The cytoskeleton also comprises molecular motors, proteins that move on actin or microtubule filaments driven by chemical energy. How the cytoskeleton in conjunction with biochemical regulatory circuits performs specific, active mechanical tasks is not understood. 
When cells attach to biological material they often biochemically recognize the binding partner. The cytoskeleton organizes accordingly
and produces a mechanical
response. 
Active cell responses such as contraction are well separated from passive rheological properties by their timescales \cite{thoumine}. 
Passive rheological cell properties have been studied with various techniques on subcellular and supercellular scale \cite{our_review}. 
From the measurements with different techniques on different eukaryotic cell types a broad relaxation spectrum arises as a common feature of passive, linear cell rheology \cite{our_review, glassy1}. 
The description of the non-linear regime remains elusive;
both stiffening
\cite{mechanotrans,prestress1,mipaper,modeltissue,monolayer}
and linear responses to large stretch
\cite{linear_saif,modeltissue,monolayer,singlecreep}
have been
observed.

\begin{figure}
\begin{center}
\includegraphics[width=0.49\textwidth]{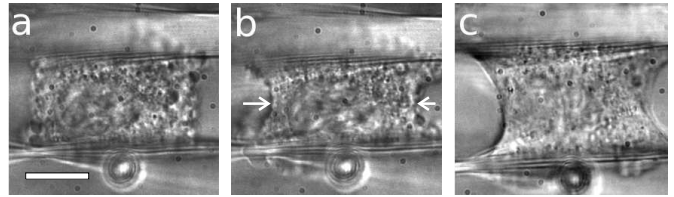}
{\black \caption{\small
\label{fig:foto}
A fibroblast adhering between two microplates.
{\bf a:} Right after contact. Bar: 10 $\mu$m.
{\bf b:} After $\sim$20 min at 35$^\circ$C,
strongly adhering cells often
adopt a concave shape.
From the apparent diameter $D_0$ (arrows)
we estimate the initial area $A_0:=\pi(D_0/2)^2$.
{\bf c:} Under large stretch.}}
\vspace{-8mm}
\end{center}
\end{figure}

In the following we present microplate rheology experiments where individual cells are stretched between two plates (Fig.\ \ref{fig:foto}). The advantage of the setup is that the cells possess a well-controlled geometry and
adhere via chosen biochemical linkers, which better define the cytoskeletal state. 
Quasi-differential cell deformations reveal an elastic stiffening response. 
The corresponding nonlinear elastic modulus depends on the cell prestress
but is independent of cell length.  
Large deformations reveal {\black an inelastic} regime
with a (counterintuitive) linear force--length relation.
Both relations  
simply superpose 
to generate the response to more complex deformations.
The cell response reduces to the integral of the differential measurement 
when the {\black inelastic} response is abolished by fixation.
Hence, in spite of the complexity of the eukaryotic cell cytoskeleton
and large cell heterogeneity, passive nonlinear cell rheology 
can be reduced to simple rules.

{\bf Experimental setup.---}
We refer to \cite{thoumine,mipaper,mitesis} for details.
A 3T3 fibroblast \cite{3T31} 
adheres between two 
fibronectin-coated glass microplates. 
One of them is flexible: its deflection
gives the perpendicular force $F$ acting on the cell.
This plate is translated by a piezoelectric actuator
controlled by a personal computer. 
The computer calculates force $F$ and 
cell length $\ell$, and adjusts the piezo position
via a proportional feedback loop
to impose a given experimental protocol.
{\black Experiments are performed at 35$^\circ$C, 
in standard medium
with Lysophosphatidic acid 50 $\mu$M (Sigma).
Cells are left to adhere for 30 min before measuring.}
All results described here are fully reproducible
for fibroblasts which adhere sufficiently strongly to
sustain pulling forces of 100 nN for several hours,
which means about 30\% of the cells in culture.

\begin{figure}[t]
\begin{center}
\includegraphics[width=0.5\textwidth]{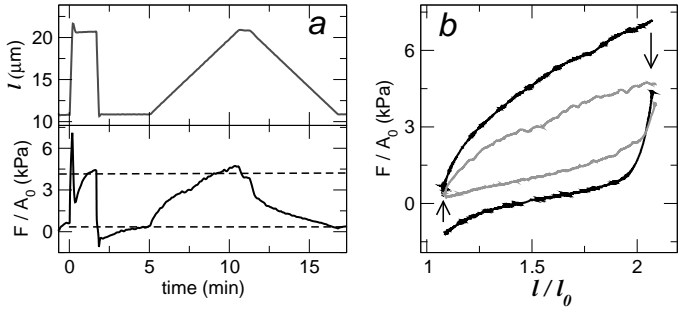}
{\black
\caption{\label{fig:large}\small
Loading and unloading at constant rate $\dot{\ell}$. 
{\bf a}: length $\ell$ and lagrangian stress $F/A_0$
(where $A_0$ is the initial area)
as a function of time.
After each ramp, $F$ stabilizes at a nonzero value $\frak{F}$ (dotted line).
{\bf b}: $F/A_0$ as a function of $\ell/\ell_0$
(where $\ell_0$
is the steady zero-force length)
during loading and
unloading.
Black curve: $\dot{\ell}=1\,\mu$m/s.
Grey curve: $\dot{\ell}=0.03\,\mu$m/s.
}
}\vspace{-2mm}
\end{center}
\end{figure}

{\bf Loading and unloading at constant rate---}
We stretch the cell by 100\%
at a constant rate $\dot{\ell}$ while measuring the force $F$
(Fig.\ \ref{fig:large}a).
The slope ${\rm d}F/{\rm d}\ell$ 
initially decreases, reaching
a constant value at an elongation $\sim 10\%$
(Fig.\ \ref{fig:large}b).
Beyond 10\% and
up to 100\% stretch, the $F(\ell)$ relation is
in most cases linear. 
After loading, 
the cell length $\ell$ is held constant for a few minutes;
the force $F$ relaxes 
to a steady non-zero value $\frak{F}$
which does not evolve faster than $\sim 1$ nN/s.
An analogous response is observed upon unloading.
The procedure is repeated 
with different rates $\dot{\ell}$
between 3 nm/s and 10 $\mu$m/s.
The asymptotic slope ${\rm d}F/{\rm d}\ell$ and
the equilibrium force $\frak{F}$ are
independent of the loading rate in the explored range.

\begin{figure}[b]
\begin{center}
\includegraphics[width=0.5\textwidth]{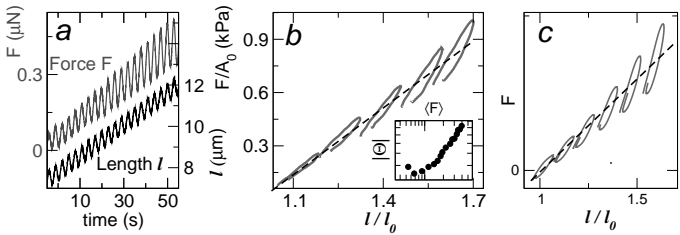}
{\black
\caption{\label{fig:ramposci}
\small
Small amplitude stiffening, large amplitude linearity.
{\bf a:} $F$ and $\ell$ as a function of time.
{\bf b:} $F/A_0$ as a function of $\ell/\ell_0$. For clarity, only a few loops
are shown. The dashed line highlights the linear
relation between the average values.
{\bf Inset:} Differential modulus $|\Theta|$ 
as a function of average force $\langle F\rangle$
for the data shown in b.  The response
to the small oscillations shows stiffening, following 
the master-relation reported in \cite{mipaper}.
{\bf c:} 
Eqs. 1--4 
 taking $\gamma=5$
and 
 $\mathcal{D}= 0.01\,\partial F/\partial\lambda$.
}}
\end{center}
\end{figure}
{\bf Small amplitude stiffening, large amplitude linearity---}
To explore small and large deformation amplitudes
simultaneously we perform
a loading ramp with superimposed harmonic oscillations,
imposing
$\ell(t)={\rm v}\,t + \Delta_\ell\sin(\omega t)$.
Fig.\ \ref{fig:ramposci} shows a typical experiment.
The response to small oscillations indeed
stiffens with increasing stress.
Yet,
the averages over an oscillation period
of the force
$\langle F\rangle$ and length $\langle\ell\rangle$
are linearly related as
inferred from the position of the loops.
Therefore, we observe both responses
\emph{simultaneously}: 
stiffening at small amplitudes, as reported in \cite{mipaper},
and linearity at large amplitudes.\linebreak

\begin{figure}[t]
\includegraphics[width=0.5\textwidth]{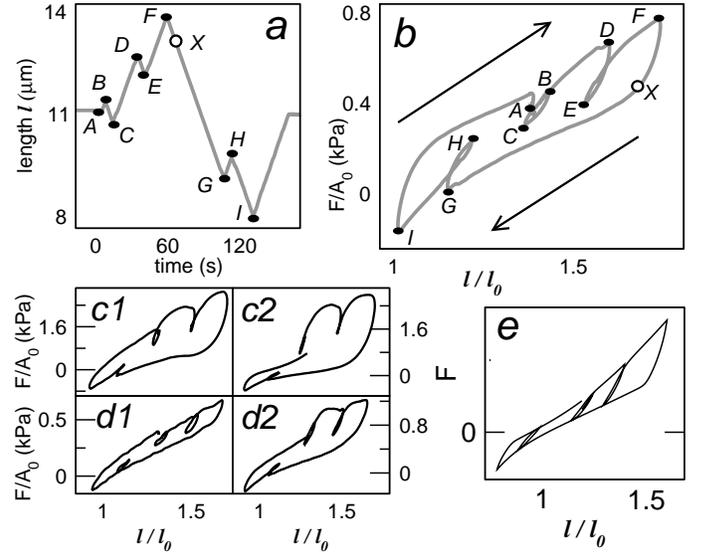}
{\black
\caption{\label{fig:demopla}\small
Elastic / inelastic behavior. 
{\bf a}: Imposed length $\ell$ as a function of time. 
{\bf b}: $F/A_0$ as a function of $\ell/\ell_0$ for a given cell.
Reversible (elastic) behavior upon direction reversal
is  observed only
close to a previous turning point,
as in C, E, H.
The response becomes irreversible (inelastic)
after a steady large deformation: at the turning points D, F, G, I,
the $F(\ell)$ curve 
does not retrace its previous path.
Between F and I the experiment is equivalent to C--F,
but in the unloading direction.
The response is seen to be equivalent,
showing the sense of deformation
to be irrelevant.
{\bf c1}: Another cell, probed at a rate $\dot{\ell}=0.1\,\mu$m/s. 
{\bf c2}: Same cell as c1 but at $\dot{\ell}=1\,\mu$m/s. 
{\bf d1}: Another cell, $0.1\,\mu$m/s. 
{\bf d2}: Same cell as d1, $1\,\mu$m/s.
{\bf e}: Prediction of the constitutive relation (Eqs.\ 1--4).}
}
\end{figure}

{\bf Small amplitude reversibility, large amplitude irreversibility---}
We study the amplitude dependence
at a constant deformation rate $\lvert\dot{\ell}\rvert$.
An essential feature of the 
protocol (Fig.\ \ref{fig:demopla}a)
are the turning points separated at various distances
in order
to study the reversibility of the response.
Similar procedures can be found in plasticity
textbooks \cite{lubliner}.
As Fig.\ \ref{fig:demopla}b shows,
the reversibility of the response upon a change
of direction is
determined \emph{by the distance to the previous
turning point}.
Where turning points are separated
by less than 10\% stretch,
the response is reversible (elastic).
More than 10\% stretch 
beyond a turning point,
the response becomes irreversible {\black (inelastic)}:
the $F(\ell)$ curve
does not retrace its path upon direction reversal. 
In this {\black inelastic} regime
the $F(\ell)$ relation is approximately linear.
Its nonzero slope leads to
\emph{a translation of the elastic region
by the {\black inelastic} deformation},
a behavior
known in plasticity
as linear kinematic (or directional) hardening \cite{Prager2,lubliner,nematnasser,VBO_krempl}.
Alternatively, the {\black inelastic} \emph{contraction} under
{\it pulling} tension between X and G in Fig.\ \ref{fig:demopla}b
is a strong Bauschinger effect 
{\black
(a decrease in yield stress upon unloading)
\cite{lubliner,constitu_crysta_plasti}.
}

{\bf Large amplitude stiffening after glutaraldehyde fixation.}---
\begin{figure}[t]
\begin{center}
\includegraphics[width=0.5\textwidth]{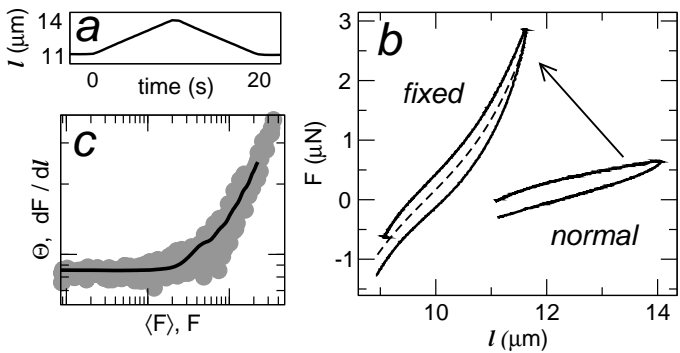}
\caption{\label{fig:glu}
\small
{\bf a}: Imposed length $\ell$ as a function of time. 
A single loading cycle with amplitude 30\% is performed.
{\bf b}: Force $F$ as a function of length $\ell$.
``Normal'': before adding glutaraldehyde.
``Fixed'': in presence of glutaraldehyde 0.1\%.
The dotted line
is a fit to Eq.\ref{eq:ela}.
{\bf c}: The derivative ${\rm d}F/{\rm d}\ell$ of the ``fixed'' curve
(black line) plotted against the differential master-relation
(gray dots, as discussed in Ref.\cite{mipaper}).
}
\end{center}
\end{figure}
We add glutaraldehyde 0.1\% 
in order to prevent slippage of cytoskeletal connections.
Loading at constant rate (Fig.\ \ref{fig:glu}a) reveals
\emph{a positive
curvature} ${\rm d}^2\!F/{\rm d}\ell^2$ (Fig.\ \ref{fig:glu}b).
The numerical derivative of the $F(\ell)$ relation 
obtained from fixed (hence dead) cells
is the same as the differential master-relation
obtained on \emph{living} fibroblasts
(Fig.\ \ref{fig:glu}c, from Ref.\ \cite{mipaper}).
The $F(\ell)$ relation after fixation closely resembles the exponential stress-stretch relations known from whole tissues \citep{fungbook,fabry}.

\begin{figure}[t]
\begin{center}
\includegraphics[width=0.5\textwidth]{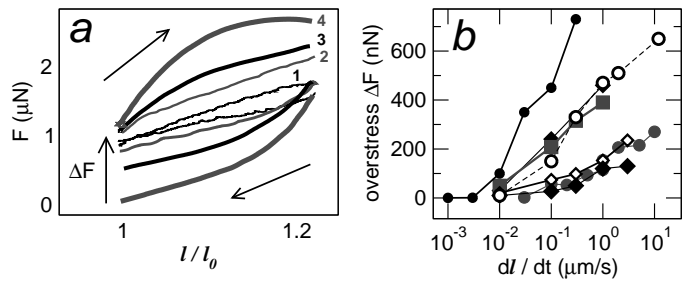}
\caption{\small\label{fig:ratl}
Loading and unloading for several rates $\dot{\ell}$.
{\bf a}: $F$ as a function of $\ell/\ell_0$ for a given cell,
where $\ell_0$ is the initial length.
Rates are: 1) 10 nm/s, 2) 30 nm/s, 3) 0.1 $\mu$m/s, 4) 1 $\mu$m/s.
The overstress $\Delta F$ is defined as the extent of relaxation
after unloading.
{\bf b}: 
{\black $\Delta F$
as a function of 
rate $\dot{\ell}$ }
for 7 different cells.
}
\end{center}
\end{figure}

{\bf Rate dependence--}
In the {\black inelastic} regime
the width of the hysteresis loops increases with
stretch rate (Fig.\ \ref{fig:ratl}a). 
To characterize this rate-dependence,
we define the overstress $\Delta F$
as the extent of force relaxation
after unloading (Fig.\ \ref{fig:ratl}a).
The overstress  $\Delta F$ behaves as $\log({\rm d}{\ell}/{\rm d}t)$,
approaching zero at a non-zero
pulling speed of about 10 nm/s
(Fig.\ref{fig:ratl}b).
{\black
Below such rates the behavior becomes active
and erratic and the overstress ill-defined.
}

{\bf Viscoplasticity--}
We now
propose a
minimal constitutive relation for fibroblasts
under uniaxial extension.
First we decompose
the measurable cell length $\ell$ into \linebreak
{\black inelastic} rest length $\frak{L}$
and elastic stretch ratio $\lambda$,
\begin{equation}
\ell = \lambda\: \frak{L} \label{eq:lple} \;.
\end{equation}
The force is a function solely
of the elastic strain,
\begin{equation}
F \propto (\lambda-1)e^{\gamma(\lambda^2+2/\lambda-3)} \label{eq:ela}\;,
\end{equation}
where for concreteness we
use exponential elasticity \cite{fungbook}, according to Fig.\ \ref{fig:glu}b, c.
As a flow rule relating
the {\black inelastic} strain rate $\dot{\frak{L}}$ to the force $F$,
we propose an exponential 
function of the overstress $F-\frak{F}$,
according to Fig.\ \ref{fig:ratl}b:
\begin{equation}
\dot{\frak{L}} = \upsilon\:
{\rm sgn}( F -\frak{F})\:
e^{ | F -\frak{F} |  / \mathcal{D}  } \label{eq:flow}\;,
\end{equation}
{\black with a vanishing dissipation
as the flow rate $\dot{\frak{L}}$
approaches $\upsilon$. 
The equilibrium force $\frak{F}$
is the essential internal variable
to describe kinematic hardening \cite{Prager2,lubliner,VBO_krempl}.
The drag force $\mathcal{D}$
sets the scale where the overstress
induces inelastic flow.
To obtain an increased dissipation at large forces
(e.g. the increase in the area of the loops in Fig.\ \ref{fig:ramposci}),
we take the drag $\mathcal{D}$ as proportional to the
nonlinear modulus, $\partial F/\partial\lambda$.
Finally, we have linear kinematic hardening:}
\begin{equation}
\dot{\frak{F}} \propto \dot{\frak{L}}\;. \label{eq:hard}
\end{equation}
{\black
This is an empiric description along the lines of modern viscoplasticity 
\cite{lubliner,constitu_crysta_plasti,VBO_krempl},
without explicit history dependencies. 
As Figs.\ \ref{fig:ramposci}c 
and \ref{fig:demopla}e show, 
it captures the essence
of the phenomenology.}
At small amplitudes, $|F-\frak{F}|\ll\mathcal{D}$,
the deformation is
essentially elastic: $\dot{\ell} \sim\dot{\lambda}$.
At large amplitudes
the overstress $|F-\frak{F}|$ approaches the drag stress
$\mathcal{D}$
and the deformation becomes increasingly {\black inelastic}:
$\dot{\ell}\sim\dot{\frak{L}}\propto
\dot{\frak{F}} \sim \dot{F}$.
{\black 
Nevertheless, this constitutive relation is
not yet a full description.
The details of the linear regime
\cite{glassy1,singlecreep} and
fluidization
at large flow rates \cite{mipaper}
still have to be incorporated to it,
whereas
active contraction and inelastic deformation
at rates below $\upsilon$,
as well as the force fluctuations
seen in Fig.\ \ref{fig:large}
may require a different approach.
}

{\bf Discussion---}
Our glutaraldehyde fixation experiments
show that stress stiffening
in fibroblasts \cite{mipaper} 
is due to the nonlinear elasticity of the
cytoskeleton, unrelated to biological
signalling or restructuring.
In agreement, very similar stiffening is
known from
biopolymer networks
\cite{nonlineargels}.
To date the precise microscopic mechanism remains unclear;
stretching \cite{nonlineargels,tissue_WLC_05kuhl} and
bending \cite{mipaper,network_kabla} of single filaments
as well as filament alignment 
\cite{entropiciscrap} have been proposed.

The energy to reach the linear {\black inelastic} regime 
in loading experiments (from Fig.\ \ref{fig:large}) is 
$\sim\:1\mu{\rm N}\,\mu{\rm m}\, \simeq 2.5\, 10^8 \,k_BT$. 
During stretch, first elastic elements must be loaded,
until they dissipate the stored energy upon bond rupture. 
Taking typical orders of magnitude
for bond energies \cite{mogilner}
and dissipation \cite{fourcade} and
a mesh size of 100 nm \cite{braybook},
over 10\%
of the actin cytoskeletal bonds 
must be ruptured to reach stationary flow.  
However, in order to observe stress stiffening
ubiquitously 
during the {\black inelastic} deformation, the actin gel must be always
above percolation threshold. 
For a gel with one bond relaxation time 
one expects a rate-independent overstress  \cite{pramana};
we observe an
exponential 
rate-overstress dependence.
In general, dissociation rates
depend on the force per bond $f$ as $\sim e^{f /f_0}$, 
with a force scale $f_0\sim$10 pN  \cite{bond_bell,bond_merkel}.
In agreement, in fibroblasts the drag $\mathcal{D}$ 
is about 100 nN,
corresponding to 1--10 pN per filament. 
Interestingly, the {\black inelastic} stretch rate where the cell
flows without hysteresis
is of the
order of $10^{-3} {\rm s}^{-1}$, 
a typical rate for active processes such as crawling
and contraction \cite{our_review, braybook}. 
Thus, 
spontaneous bond dissociation
may be what limits
active phenomena
to long timescales \cite{thoumine}.

{\black
A living cell can neither be purely elastic,
nor possess a yield stress within the physiological ``working range''.
Kinematic hardening viscoplasticity 
can be understood as a consequence of these conditions.}
Combined with a sharply rising rate-overstress
dependence, it
prevents cell breakage in our large deformation
experiments:
a cell portion under increased stress will
readily flow and 
increase its equilibrium stress
to reach
a stable situation.
Rather than break at a given spot, the cell prolongs homogeneously
along its length.
This homogeneous deformation may be 
behind the robust linearity of the kinematic hardening response,
since integration of a constant magnitude 
along the cell length 
naturally gives a linear length-dependence.
However, identification of the precise molecular mechanisms 
behind this unusual behavior in a soft system is a task for the future.
{\black
At least, one can rule out a trivial interpretation in terms of
a Hookean spring element (in form of intermediate filaments, for example) 
in parallel with a stiffening viscoelastic liquid:
since the liquid cannot sustain an average stress,
a non-mechanical coupling between the two elements
is needed for the cell to stress-stiffen. Thus, in a mechanical interpretation
hardening and stiffening
must originate in one and the same mechanical element.
If intermediate filaments \cite{braybook} play a role,
they must be interconnected with actin into a single network.}
This scenario reminds of
composite alloys \cite{constitu_crysta_plasti}
and granular materials \cite{nematnasser}, where kinematic hardening
arises as the {\black inelastic} flow induces alterations of directional nature
to the microstructure.

Summarizing, we have shown that cell mechanical properties in uniaxial stretching experiments can be thoroughly described by the superposition of two simple relations: 
exponential elasticity, and
viscoplasticity with linear kinematic hardening. 
Given the cytoskeletal complexity, this is unexpected. A complete picture of passive cell rheology spanning from molecular details to a simple phenomenological description and straightforward theoretical concepts seems in close reach.

{\black We thank P. A. Pullarkat for his invaluable advice,
and K. Kroy for inspiring discussions and support.}
This work has been funded by the Universit\"at Bayreuth.


\begin{thebibliography}{29}
\bibitem[{fab()}]{fabry}
\bibinfo{note}{Exponential elasticity can also be revealed by microrheological
  techniques. B. Fabry, personal communication}.
\end{thebibliography}
\end{document}